# Colloidal Clusters as models for chiral active micromotors

Bipul Biswas[1] and Manasa Kandula[1, *]

[1]Physics Department, UMass Amherst

Abstract

Circular swimmers with tunable orbit radius and chirality are gaining attention due to their potential to illustrate novel collective phases in simulations and synthetic and biological active matter. Here, we present a facile experimental strategy for fabricating active rotors using chemically cross-linked clusters of Janus colloids. Janus clusters are propelled by induced charge electrophoresis in an alternating electric field. We demonstrate capillary-assisted assembly as a feasible path toward expanding the fabrication process to get large amounts of uniform circular clusters. Systematic studies of the Janus clusters reveal circular motion with tunable angular velocity, orbit radius, and chirality and a relation between the radius of gyration of the cluster and their rotational dynamics. Importantly, clusters with uniform azimuthal angles behave distinctly exhibiting larger orbit radii, while those with random angles exhibit higher angular velocities and smaller radii. We also validate the kinetic model for clusters beyond dimers. Collectively, our studies highlight that Janus clusters as promising candidates as controlled circular rotors with tunable properties and, hence in the future, will facilitate studies on the collective behaviors of synthetic chiral rotors.

# Introduction

Circular swimmers with tunable orbit radii and chirality are increasingly being recognized for their potential in both theoretical and applied sciences. These swimmers are particularly intriguing due to their prevalence in natural systems and their ability to form novel collective behaviors.[1–5] Theoretical and simulation studies have extensively explored the phase diagrams of these circular swimmers, uncovering exotic collective states such as disordered hyperuniformity in the phase space of trajectory radius and particle number density.[1,6–12] In practical applications, circular swimmers offer significant advantages in microfluidics at low Reynolds numbers. They generate chaotic flow, which is valuable for cargo transport and biomedical applications.[13,14] Over recent years, advances have been made in fabricating synthetic colloidal micromotors that exhibit a range of motile trajectories, including run-and-tumble, active Brownian motion, spinning, circular trajectories, artificial flagella and gears.[6,15–18] Despite these advances, there remains a limited number of colloidal systems that have tunable circular trajectories but also switch their chirality and/or transition to linear motion on demand.[19–22]

In this work, we propose a model system of micromotors that propel in circular orbits (as opposed to spinning) based on colloidal clusters. Literature suggests that colloidal clusters are promising models for circular swimmers with tunable trajectories.[9,23–25] For instance, self-assembled clusters of Janus colloids driven by diffusophoresis or induced charge electrophoresis (ICEP) have been demonstrated to follow circular paths.[16,26,27] Nevertheless, there is a notable lack of systematic studies on the circular dynamics of active colloidal clusters, particularly regarding control over angular velocity, orbit radius, chirality, and the large-scale production necessary for collective studies. To address these gaps, we focus on gaining a comprehensive understanding of the self-propulsion mechanisms of Janus colloid clusters in alternating electric fields and developing strategies for large-scale synthesis to facilitate collective behavior studies.

For a Janus colloid under an external AC field, propulsion and the direction of force are directly related to the orientation of the Janus hemisphere. Consequently, the relative orientation of particles within a cluster significantly influences the nature of its trajectory.[28,29] Experimental studies have highlighted the importance of relative orientations in Janus dimers, with ICEP-driven and catalytically driven clusters showing variations in behavior based on these orientations.[16,26] For instance, research by Boymelgreen et al. and Ebbens et al. demonstrated that catalytic Janus colloid particles tend to self-organize into dimers with preferred relative orientations, constrained by particle interactions.[16,30] However, chemically cross-linked clusters, unlike self-organized ones, can overcome these limitations and explore a broader phase space. Therefore, fabricating cross-linked clusters is essential for comprehensive studies of active cluster dynamics. This approach will allow us to investigate how cluster shape and size affect self-propulsion dynamics beyond the dimers and not constrained by particle interactions.

Here, we investigate how the shape, size, and orientations of Janus particles affect the dynamics of chemically cross-linked rigid active clusters. To identify the key parameters influencing cluster dynamics, we fabricate clusters composed of metal-dielectric (Titanium-Polystyrene) Janus colloids through an ice-templating method.[31–33] We then employ capillary-assisted assembly as a strategy to produce a large number of clusters with uniform shape and size.[34] Our dynamics studies of clusters in an orthogonal alternating electric field (AC electric field) reveal that trimers and tetramers exhibit predominantly circular motion, unlike dimers. This behavior is attributed to the non-uniform electro-hydrodynamic forces generating torque. The observed dynamics of trimers are consistent with the predictions of Squires and Bazant, indicating that induced charge electrophoresis is responsible for the circular motion.[35] Crucially, our chemically cross-linked clusters allow for in-situ tuning of angular velocity, orbit radius, and chirality without disassembling the clusters. Interestingly, we find that clusters with all Janus colloids oriented at the same azimuthal angle (orthogonal clusters) have a significantly larger orbit radius compared to clusters with randomly distributed polar and azimuthal angles (planar clusters), even when the geometry (shape and size) is identical. Data analysis based on the radius of gyration shows that the experimentally measured angular velocity scales linearly with the velocity predicted by the kinetic model, validating the model's applicability to trimers and tetramers. Our findings highlight that active clusters are excellent candidates for circular rotors, with orthogonal clusters providing effective control over orbit radius and planar clusters ideal for tuning angular velocity. We anticipate that our results, combined with scalable synthesis methods, will advance the study of collective behaviors of synthetic circular micro-motors.

# Results

## Fabrication of clusters:

We fabricate chemically cross-linked clusters to explore a wide range of geometric parameters, including cluster shape, size, and the relative orientations of Janus colloids. Our initial approach involves using ice templating to synthesize clusters with various geometries and random orientations (Figure 1A). Metal-dielectric Janus colloids are created by sputtering titanium onto particles, following established protocols.[36] We then modify the surface of these particles with Poly(N-isopropylacryla-mide co-allylamine) microgel[37] (PNIPAM-NH$_2$) for chemical cross-linking. Dilute suspensions of these functionalized particles are mixed with polyethylene glycol di-epoxide (PEG-di-epoxide) solutions. This mixture is frozen at –20°C, causing water to solidify into ice. During this process, the colloids assemble into clusters at the ice grain boundaries, and PEG-di-epoxide facilitates the cross-linking of individual spherical colloids into planar clusters.[31–33] We extract clusters by melting the ice. We optimize the starting volume fraction to obtain clusters with fewer than five particles predominantly. Our experiments reveal that the ice crystallization method yields clusters with random Janus hemisphere orientations (Figure 1C), referred to as *"planar clusters"* in the rest of the manuscript. To create clusters where all particles have the same azimuthal orientation of the Janus hemisphere, we first form passive colloid clusters and then deposit metal via sputtering to create Janus particles. These are referred to as *"orthogonal clusters"* henceforth (Figure 1D).

To achieve mass production of clusters with specific shapes and sizes, we combine the cross-linking protocol with capillary-assisted assembly (CAA).[34,38] PNIPAM-NH$_2$ coated colloids are dispersed in a 0.02 wt% Triton-water mixture, and a drop of this dispersion is placed on PDMS-patterned surfaces. Capillary forces and local confinement from the structured cavities on the template control particle placement (Figure 1E). Using step-wise CAA, we deposit the desired number of particles into the cavities (Figure 1F).[34] After washing the surface with DI water, we cross-link the particles by freezing and extract the clusters by melting the ice, following the same protocol as the ice crystallization method. Figure 1G shows an optical microscope image of multiple trimers prepared by CAA. This method allows precise control over the shape and size of the clusters, with the number of clusters determined by the mask area.

## Trajectories of active clusters:

We investigate the active motion of our clusters by subjecting them to alternating electric fields (AC electric field) perpendicular to the imaging plane (Figure 1B). We apply AC fields ranging from 0.012 to 0.8 V/μm and frequencies from 2 kHz to 1.5 MHz to explore both induced charge electrophoresis (ICEP) and dielectrophoresis (sDEP) regimes. Using optical microscopy, we capture time-series images and analyze them with custom-developed algorithms to quantify cluster shapes and dynamics using MATLAB. When a spherical metal-dielectric Janus colloid is exposed to an AC electric field, it reorients due to the competition between thermal fluctuations, gravitational potential, and field-induced polarizability. The difference in polarizability between the two hemispheres causes asymmetric fluid flow, resulting in local forces and propulsion.[28,29] Consequently, the steady-state orientation of the clusters depends on their shape and the orientation of the particles within, leading to asymmetric force distribution. We observe that orthogonal clusters undergo significant reorientation and align perpendicular to the imaging plane in the direction of the electric field (Figure 1D). Therefore, we refer to these as orthogonal clusters. In contrast, planar clusters exhibit moderate reorientation but remain in the plane parallel to the electrodes (i.e., the imaging plane, Figure 1C).

Reoriented clusters self-propel due to the asymmetric flow field around them. Unlike spherical particles, this asymmetric flow generates both a force and a torque, causing trimers and tetramers to predominantly move in circular orbits, irrespective of their relative orientations (see SI Videos 1 & 2). A cluster will exhibit pure translational motion only if all particles are oriented in the same plane, which is rare. Figure 2 illustrates representative circular trajectories and their corresponding mean squared displacements, showing oscillations at long times for trimers and tetramers subjected to an AC electric field of 0.048 V/μm and 3 kHz. We observe that the orbit diameter ($D_{Orbit}$) depends on the cluster shape, size, and the relative orientations of the Janus colloids (Figure 2). Notably, compact trimers and tetramers with orthogonal orientations exhibit a larger $D_{Orbit}$ compared to planar clusters under the same AC field conditions (Figure 2D & G, E & H). In contrast, linear clusters with orthogonal orientations predominantly display translational motion, while planar ones continue to exhibit circular trajectories.

## AC electric field as a tool to control orbit dynamics in-situ:

Having established that the circular orbit of an active cluster depends on its geometry and the orientation of Janus colloids, we now demonstrate how to engineer the cluster's trajectory by tuning the AC electric field strength (E) and frequency (f). Chemical cross-linking is crucial for this tuning, as without it, inter-particle interactions such as dipolar repulsion would prevent effective trajectory manipulation. To compare across trajectories, we estimate the angular velocity ($\omega_T$) and $D_{Orbit}$. We plot corrected Ferret's angle ($\varphi$) as a function of time (Figure-SI 1 & S2) and extract $\omega_T$ from the linear fit of the data points (Figure-SI 2C). We find the mean and standard deviation of $D_{Orbit}$ by estimating the displacement difference between two successive minima and maxima corresponding to X and Y coordinates of the circular trajectories (Figure-SI 2).

For any given cluster, we observe that $\omega_T$ increases linearly with increasing $E^2$ for all planer clusters while the slope depends on the relative orientation of the Janus colloids in that cluster (Figure 3A). The observed increase in $\omega_T$ with $E^2$ indicates an increase in the propulsion speed of individual Janus colloids manifested as increasing torque for clusters. The $E^2$ dependence is consistent with the translational velocity ($v$) of spherical Janus colloids proportional to $E^2$, confirming that the propulsion mechanism is ICEP at these values of E and f. Along with $\omega_T$, we find that the $D_{Orbit}$ also increases moderately with increasing $E^2$ (Figure-3B). For orthogonal clusters, we see a more pronounced increase in $D_{Orbit}$ while $\omega_T$ does not change linearly (Figure-3A). The deviation from linearity is primarily due to the cluster's tilt angle relative to the XY-plane. With increasing field, the dipoles align strongly along the field direction making the cluster orient fully perpendicular to the plane. At high field strengths (E>0.048V$\mu m^{-1}$) we observe a translation motion occasionally when the cluster is fully aligned along the electric field.

Next, we investigate the dependence of $\omega_T$ on the frequency of the AC field. It is known that a spherical Janus colloid reverses its direction of motion beyond a critical applied frequency of the AC field.[20] This direction reversal in individual colloids is likely to affect the chirality of clusters as they are propelling in circular orbits. To confirm this, we studied dynamics of various active clusters over a range of frequencies for a fixed E = 0.048V$\mu m^{-1}$ (Figure-3C). We find that regardless of the cluster geometry $\omega_T$ decreases with $f$. Beyond a critical frequency, typically beyond 200kHz, we notice a reversal in the chirality (Figure-3C). This frequency coincides with the proposed mechanism change from ICEP to sEDP. It is important to note that chemical cross-linking is crucial for observing frequency-dependent behavior. Without chemical crosslinking the cluster will disassemble due to increased repulsion between hemispheres of same species with frequency of the applied field increases.[27,39]

We further exploit the frequency and field dependence to engineer the trajectories of the active clusters shown in the three examples in Figure 2D-E. First, we demonstrate trajectory switching between circular to linear motion (Figure-3D). At E=0.048V$\mu m^{-1}$, f=3kHz the cluster moves clockwise with $\omega_T = 0.42$ rad/s. As the frequency increases to 260kHz, the motion switches to linear. Increasing the frequency further to f=600kHz transforms the linear trajectory into a counterclockwise circular motion with $\omega_T = 0.12\ rad/s$. Second, we show that a spiral trajectory can be achieved by logarithmically/linearly increasing the frequency for a fixed applied voltage (Figure-3E, Figure SI 7A and SI-Video-4). Finally, chirality reversal can be engineered by instantaneously changing the frequency of the applied field from f=3kHz to f=600kHz (Figure-3F, Figure SI 7D and SI-Video-5).

## Kinetic Model for describing active clusters:

Thus far, we have concentrated on quantifying the behavior of single clusters and exploring strategies to control their trajectories. We now extend the kinetic model discussed in reference 16 to trimers and tetramers to predict angular velocity and chirality from static images, relying solely on the direction of forces on individual Janus colloids within the cluster. For each Janus colloid, we determine the direction of force as the vector pointing from the center of the metal hemisphere (dark hemisphere) to the center of the dielectric hemisphere (bright hemisphere) under ICEP regimes (see Figures SI 3 and S4).

For a planar cluster, Figure-4A illustrates the force direction based of the orientation of Janus colloids assuming all Janus colloids move with the same speed $V_{JP}$. From single particle analysis we find the average speed to be 1.78±0.28 μms$^{-1}$ (see details in S4). Figure-4A clearly suggests that the net force induces a torque on the active cluster. Using kinetic model, we calculate the angular velocity $\omega_I$ as (see SI)

$$\omega_I = \sum_i^N \hat{k}(x_i \sin\alpha_i - y_i \cos\alpha_i)|V_{JP}|/R^2$$

Figure-4B shows a comparison between the direction of motion from the trajectory (left) and the instantaneous direction computed by the kinetic model, demonstrating good agreement. Figure SI 5 confirms that this estimate is robust against variations in image processing details.

For orthogonal clusters (Figure-4C) we must also consider the tilt of the cluster relative to the z-axis, as indicated by our video microscopy experiments (Figure SI 9). Comparing the trajectories of tilted versus non-tilted clusters confirms that only the tilted clusters display circular trajectories, while non-tilted clusters exhibit translational motion, consistent with recent theoretical studies that highlight the importance of tilt for circular trajectories (Figure-4D). Additionally, we observe that linearly propelling clusters (those without tilt) appear only at very high electric fields and sporadically, suggesting this occurs only when dipoles align perfectly along the field direction. Indirectly the kinetic model confirm that orthogonal clusters are titled as the $\omega_I$ is zero without taking the tilt angle into accounts. The Measurement of the tilt angle (ϕ) experimentally is challenging and is beyond the scope of the current study (See details in SI and Figure SI 9).

To quantitatively compare the angular velocity estimates from the kinetic model with experimental data for over 50 different clusters, we categorize the clusters based on the ratio $R_g/R$, where $R_g$ represents the radius of gyration and R is the radius of a single Janus colloid. Clusters are classified into five categories: two-particle (2), compact three-particle (3C), open three-particle (3O), compact four-particle (4C), and open four-particle (4O) rotors (see Figure SI 6). The validity of the kinetic model for planar clusters is assessed by plotting $\omega_T$ versus $\omega_I$ angular velocity. Notably, for each category of the cluster $\omega_T$ versus linearly with $\omega_I$ for a fixed E=0.048Vμm$^{-1}$ and f=3kHz (Figure 4A). The slope of the linear fit varies with cluster category, and the fits do not pass through the origin, which we attribute to changes in drag caused by shape, electric field, and density differences between particles and solvent[16]. By normalizing $\omega_I$ and $\omega_T$ with $\omega_{Imax}$ and $\omega_{Tmax}$ (slope×$\omega_{Imax}$) respectively we achieve data collapse indicating that the effects of geometric parameters are consistent within clusters of a given shape, $R_g/R$ (Figure-4B). Here $\omega_{Imax}$ is the maximum angular estimated by the kinetic model for a cluster of a specific category (details in Figure SI 10).

Since the orbit radius is a key parameter for realizing different collective phases, we examine the trends in orbit radius for both planar and orthogonal clusters. When the angular velocity is high, the cluster tends to spin similarly to what is observed in Quincke rollers and colloidal rollers. To achieve circular trajectories rather than spinning, it is crucial to use clusters such as trimers and tetramers with compact shapes and orthogonal clusters (Figure-4C). Plotting the diameter of the circular orbits, $D_{Orbit}$, as a function of $R_g$ interestingly reveals that the diameter can be tuned by adjusting the shape and size of the clusters. Importantly, we find that orthogonal clusters exhibit trajectories with larger diameters and show a greater dependence on cluster shape (see Figure 4D). In contrast, dimers of orthogonal clusters exhibit only translational motion, highlighting the necessity of using clusters larger than dimers to achieve circular trajectories. Overall, Figure-4 demonstrates that clusters are an ideal system for developing circularly swimming micromotors. By manipulating the electric field and cluster geometry, we can control the orbit diameter, which is crucial for investigating collective phase transitions in these systems.

## Conclusions:

In conclusion, our studies provide a new facile strategy to make circular active colloids. Our planar and non-planar structures offer an easy strategy to make active rotors with the desired angular speed and orbit diameter. While controlling the orientation of every Janus colloid and producing rotors in bulk remains challenging, our studies show that fabricating clusters of a given shape and size ($R_g/R$) will serve as experimental models to study collective behavior in circular and chiral swimmers. The capillary assisted assembly puts a path forward to produce large numbers of homogeneous clusters of a certain shape and size thereby enabling the study of collective motion. We show that the kinetic model can be extended to trimer and tetramers in plane and orthogonal. We expect our study to prompt experimental studies to use clusters as a strategy to make circular swimmers with engineered trajectories and also use them for collective phenomenon and pattern formation studies.

## Acknowledgements

MK acknowledges financial support from the University of Massachusetts Amherst for starting faculty support. BB acknowledges Dr. Michael Wang from the PSE Department at UMass for his valuable discussions.

## Author contributions

BB performed experiments. BB and MK analyzed the data and wrote the manuscripts. Research was supervised by MK.

## Corresponding author

Department of Physics, University of Massachusetts Amherst, MA-01002, USA; Email-*hkandula@umass.edu.*

Figures:

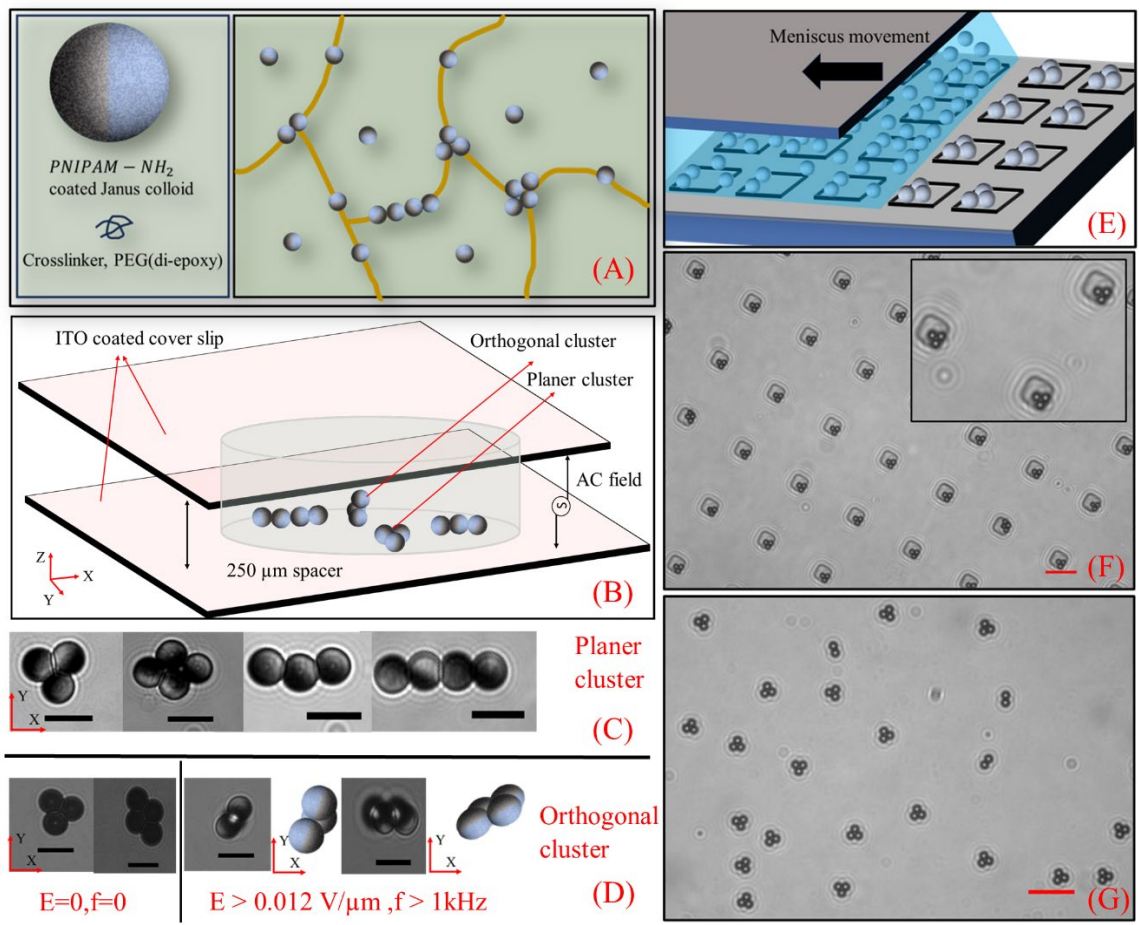

***Figure 1: Experimental setup and colloidal clusters:*** *(A) PNIPAM-NH$_2$ micro-gel coated Janus colloid and cross-linker (PEG diepoxy) are mixed in aqueous dispersion, after that the dispersion is cooled at $-20^oC$, PNIPAM-NH$_2$ coated Janus colloids and cross-linker accumulates along the boundary of the ice crystal. PNIPAM-NH$_2$ forms a crosslinked mess and Janus particles are linked into clusters with different radius of gyrations. Ice crystal is then thawed, and we harvest the dispersion containing clusters. (B) An electric field setup consists of a 250 µm thick silicon spacer with a circular hole (sample chamber), which is used to separate two ITO-coated cover slips. The electric field is applied perpendicular to the ITO surfaces. (C)Representative snapshots of planer clusters, where there is a minimum reorientation after applied field (D) Representative snapshots orthogonal clusters where there is a significant reorientation are seen after applied field. (E) Schematic diagram of Capillary assisted assembly (CAA) to make uniform size and shaped clusters. (F) Three particle clusters are on the PDMS template (G) After harvesting the clusters from PDMS surface by ice lifting, we show a snapshot of dense clusters with same size and shape.*

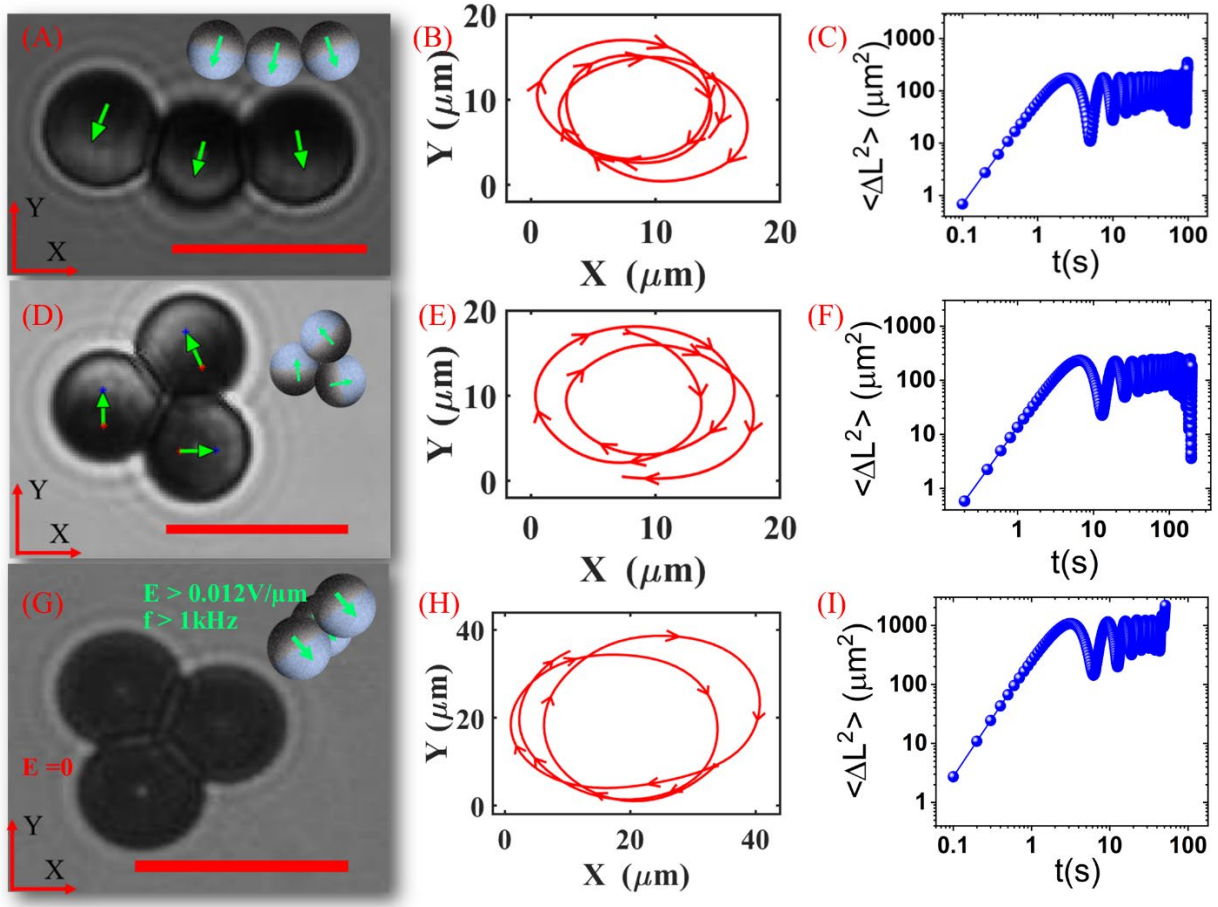

***Figure 2: Circular motion***: *(A) Representative planer cluster of 3 particle open structure: all the three velocity vectors (green arrows) for individual colloids are shown (B)-(C) Trajectories of the rotor and the corresponding MSD are shown respectively. (D-F) Representative planer cluster of 3 particle compact structure: trajectories for the cluster and the MSD of the trajectories are shown. (G-I) Representative orthogonal cluster of 3 particle compact structure: its trajectories and MSD are shown. All the scale bar is 5µm. Time for all the trajectories that are shown in (B, E&H) for three clusters are same.*

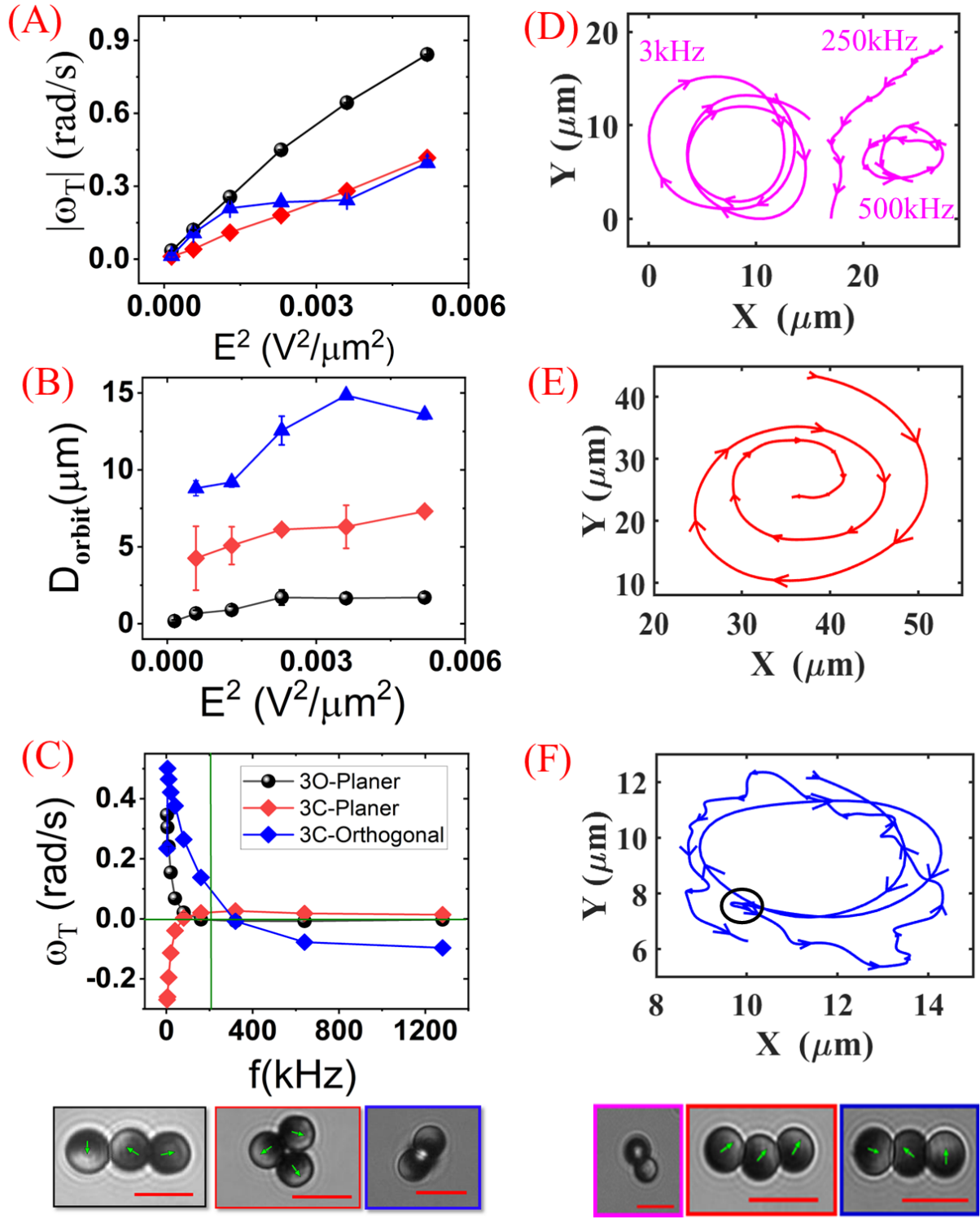

*Figure 3: Field and frequency dependent dynamics:* (A) Variations of angular velocity ($\omega_T$) are plotted as a function of applied field square ($E^2$) for three different types of clusters (planer cluster- 3 particles open structure & compact cluster and orthogonal cluster- 3 particles compact cluster). Slopes of the line are different from each other, and it depends on the orientations of the individual colloids, shape, size and electro-viscous drug of the clusters. For orthogonal cluster: we observe that the variations of angular velocity with $E^2$ are not linear, initially it increases rapidly with $E^2$ and after that there is very slow increment of the angular velocity with $E^2$. These variations come due to the reorientations of cluster along the field and that changes the electro-viscous drug. (B) Variations of orbit diameter of the circular trajectories are platted as function of applied field square for the same three clusters, and it also increases with field. (C) Variations of angular velocity of three different clusters: planner open & compact cluster, and for an orthogonal compact cluster as a function of frequency are shown. Chirality of the angular velocity changes at above the critical frequency 200kHz. (D) We have shown the trajectories clusters (shown below the trajectories) across three distinct frequency ranges where ICEP predominates at 3kHz. Within the crossover frequency range (100kHz to 300kHz), the clusters display a slow yet directional motion. Furthermore, at higher frequencies, as illustrated at 500kHz, we note the reversal motion of the clusters attributed to sDEP. (E) Cluster shows spiral motion when the applied frequency is changed as logarithmic function. (F) Reversing rotation is shown when we increase the frequency as a step function. Initially the frequency was set at 3 kHz, and we change the frequency to 500 kHz and within a few seconds cluster turns its directions of rotational motion, trajectories are marked with black circle indicates the time when frequency was change to 500 kHz and its rotational direction changes. We have shown the cluster and the step-function frequency with time are just below the trajectories. Applied field for all the cases (D-F) is $0.048 V\mu m^{-1}$.

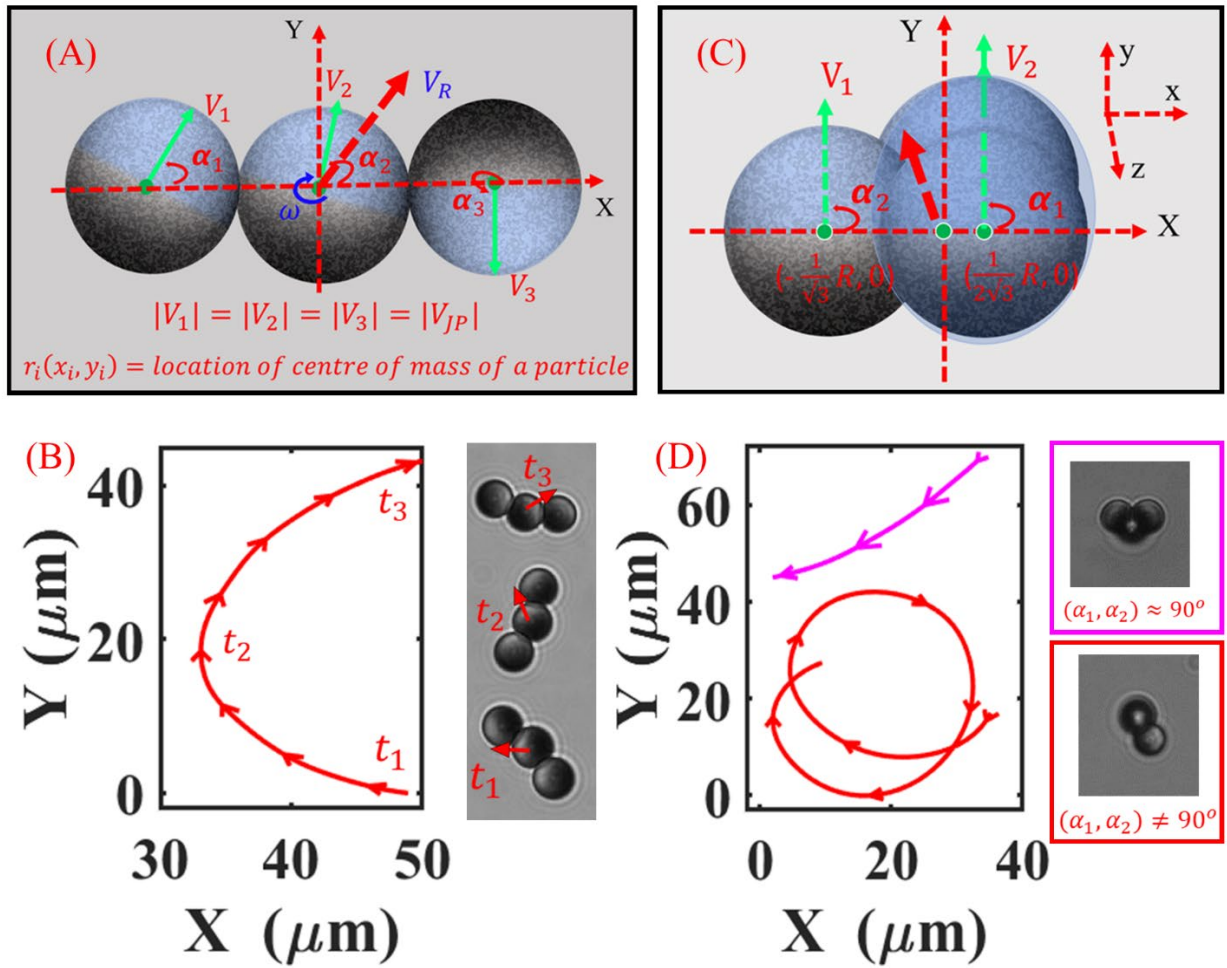

***Figure 4: Kinetic model:*** *(A) Schematics of a 3O-planer cluster, $V_1, V_2,$ and $V_3$ are the direction of velocity vectors for the Janus colloids. $V_R$ is the resultant vector of the three velocity vectors. The angles $\alpha_1, \alpha_2,$ and $\alpha_3$ are the angles of the vectors that are made from the x-axis. We assume the magnitude of the velocity vector for each Janus colloids is same and it's equal to single Janus particles speed $|\vec{V_{JP}}|$. (B) Trajectories of the active cluster follow the resultant velocity vector direction of the cluster. (C) Schematics of a 3C-orthogonal cluster, where orientation of all the particles are in the same direction. (D) Trajectory of an 3C-orthogonal cluster under AC field when the angles are equal to $90^0$ (magenta color). The trajectory of the same cluster under AC field when the angles are not equal to $90^o$ (red color).*

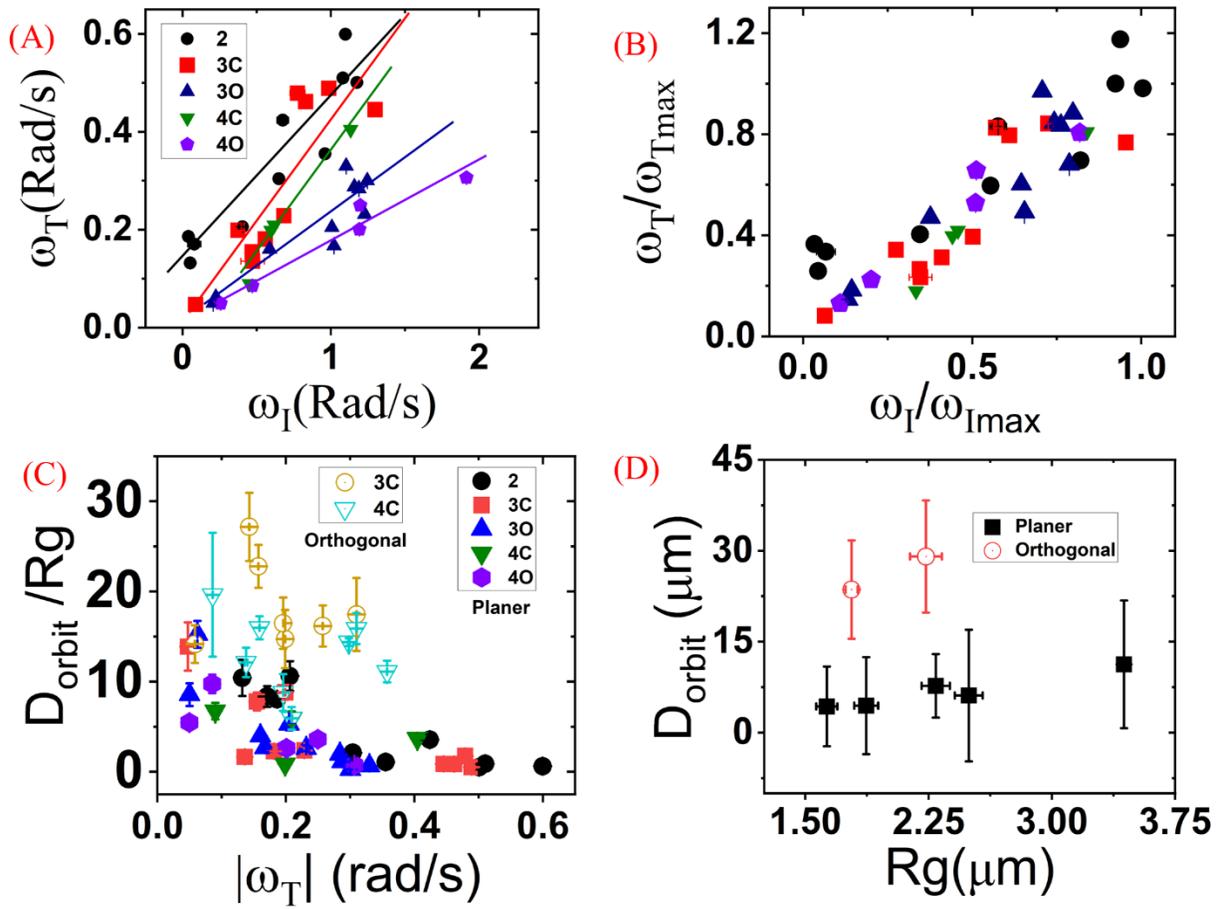

***Figure 5: Angular velocity and Orbit diameter:*** *(A) We plot the angular velocity ($\omega_T$) of the rotors with the angular velocity that we estimated form the image ($\omega_I$) for different shape and size of the clusters. Lines are the fitting of the data points: black lines are for the fitting of the black points and so on. (B) We plot the normalized $\omega_T$ and $\omega_I$ with $\omega_{Tmax}$ and $\omega_{Imax}$, respectively and we observe the all the data indeed collapses all the angular velocities onto a single straight line, here $\omega_{Imax}$, is the maximum angular velocity a cluster can have for a certain type. (C) Normalized Orbit diameter with the Rg of the clusters are plotter with absolute angular velocity that we estimate form the trajectories of the cluster. For each type of cluster, for a fixed frequency and field, the normalized orbit diameter reduces with $\omega_T$ (D) We plot the mean orbit diameter as a function of Rg of the clusters. The diameter of the planer clusters doesn't change with size and shape of the cluster, but orthogonal clusters show a large trajectory radius and furthermore show a significant change with the shape of the cluster.*

# References:

1. Fernández-Medina, M., Ramos-Docampo, M. A., Hovorka, O., Salgueiriño, V. & Städler, B. Recent Advances in Nano- and Micromotors. *Adv. Funct. Mater.* **30**, 1908283 (2020).

2. Lauga, E., DiLuzio, W. R., Whitesides, G. M. & Stone, H. A. Swimming in Circles: Motion of Bacteria near Solid Boundaries. *Biophys. J.* **90**, 400–412 (2006).

3. Nakane, D., Odaka, S., Suzuki, K. & Nishizaka, T. Large-Scale Vortices with Dynamic Rotation Emerged from Monolayer Collective Motion of Gliding Flavobacteria. *J. Bacteriol.* **203**, 10.1128/jb.00073-21 (2021).

4. Rabani, A., Ariel, G. & Be'er, A. Collective Motion of Spherical Bacteria. *PLOS ONE* **8**, e83760 (2013).

5. Kümmel, F. *et al.* Circular Motion of Asymmetric Self-Propelling Particles. *Phys. Rev. Lett.* **110**, 198302 (2013).

6. Huang, M., Hu, W., Yang, S., Liu, Q. X. & Zhang, H. P. Circular swimming motility and disordered hyperuniform state in an algae system. *Proc. Natl. Acad. Sci. U. S. A.* **118**, 1–8 (2021).

7. Löwen, H. Chirality in microswimmer motion: From circle swimmers to active turbulence. *Eur. Phys. J. Spec. Top.* **225**, 2319–2331 (2016).

8. Hoell, C., Löwen, H. & Menzel, A. M. Dynamical density functional theory for circle swimmers. *New J. Phys.* **19**, 125004 (2017).

9. Liebchen, B. & Levis, D. Collective Behavior of Chiral Active Matter: Pattern Formation and Enhanced Flocking. *Phys. Rev. Lett.* **119**, 1–5 (2017).

10. Yang, Y., Qiu, F. & Gompper, G. Self-organized vortices of circling self-propelled particles and curved active flagella. *Phys. Rev. E* **89**, 012720 (2014).

11. Ni, S., Leemann, J., Buttinoni, I., Isa, L. & Wolf, H. Programmable colloidal molecules from sequential capillarity-assisted particle assembly. *Sci. Adv.* **2**, e1501779 (2016).

12. Liu, P. *et al.* Oscillating collective motion of active rotors in confinement. *Proc. Natl. Acad. Sci.* **117**, 11901–11907 (2020).

13. Biswal, S. L. & Gast, A. P. Micromixing with Linked Chains of Paramagnetic Particles. *Anal. Chem.* **76**, 6448–6455 (2004).

14. Wang, J. & Manesh, K. M. Motion Control at the Nanoscale. *Small* **6**, 338–345 (2010).

15. Lozano, C., Gomez-Solano, J. R. & Bechinger, C. Run-and-tumble-like motion of active colloids in viscoelastic media. *New J. Phys.* **20**, 015008 (2018).

16. Boymelgreen, A., Yossifon, G., Park, S. & Miloh, T. Spinning Janus doublets driven in uniform ac electric fields. *Phys. Rev. E* **89**, 1–5 (2014).

17. Sokolov, A., Apodaca, M. M., Grzybowski, B. A. & Aranson, I. S. Swimming bacteria power microscopic gears. *Proc. Natl. Acad. Sci.* **107**, 969–974 (2010).

18. Electric-field–induced assembly and propulsion of chiral colloidal clusters | PNAS. https://www.pnas.org/doi/abs/10.1073/pnas.1502141112.

19. Shelke, Y., Srinivasan, N. R., Thampi, S. P. & Mani, E. Transition from Linear to Circular Motion in Active Spherical-Cap Colloids. *Langmuir* **35**, 4718–4725 (2019).

20. Huo, X., Wu, Y., Boymelgreen, A. & Yossifon, G. Analysis of Cargo Loading Modes and Capacity of an Electrically-Powered Active Carrier. *Langmuir* **36**, 6963–6970 (2020).

21. Lin, Z. *et al.* Magnetically Actuated Peanut Colloid Motors for Cell Manipulation and Patterning. *ACS Nano* **12**, 2539–2545 (2018).

22. Gao, Y. *et al.* Dynamic Colloidal Molecules Maneuvered by Light-Controlled Janus Micromotors. *ACS Appl. Mater. Interfaces* **9**, 22704–22712 (2017).

23. Lei, T., Zhao, C., Yan, R. & Zhao, N. Collective behavior of chiral active particles with anisotropic interactions in a confined space. *Soft Matter* **19**, 1312–1329 (2023).

24. Negi, A., Beppu, K. & Maeda, Y. T. Geometry-induced dynamics of confined chiral active matter. *Phys. Rev. Res.* **5**, 023196 (2023).

25. Bricard, A., Caussin, J.-B., Desreumaux, N., Dauchot, O. & Bartolo, D. Emergence of macroscopic directed motion in populations of motile colloids. *Nature* **503**, 95–98 (2013).

26. Johnson, J. N. *et al.* Dynamic stabilization of Janus sphere trans-dimers. *Phys. Rev. E* **95**, 1–7 (2017).

27. Zhang, J., Yan, J. & Granick, S. Directed Self-Assembly Pathways of Active Colloidal Clusters. *Angew. Chem. - Int. Ed.* **55**, 5166–5169 (2016).

28. Gangwal, S., Cayre, O. J., Bazant, M. Z. & Velev, O. D. Induced-Charge Electrophoresis of Metallodielectric Particles. *Phys. Rev. Lett.* **100**, 058302 (2008).

29. Boymelgreen, A. M. & Miloh, T. Induced-charge electrophoresis of uncharged dielectric spherical Janus particles. *ELECTROPHORESIS* **33**, 870–879 (2012).

30. Ebbens, S., Jones, R. A. L., Ryan, A. J., Golestanian, R. & Howse, J. R. Self-assembled autonomous runners and tumblers. *Phys. Rev. E - Stat. Nonlinear Soft Matter Phys.* **82**, 6–9 (2010).

31. Kumaraswamy, G., Biswas, B. & Kumar Choudhury, C. Colloidal assembly by ice templating. *Faraday Discuss.* **186**, 61–76 (2016).

32. Biswas, B. *et al.* Linking Catalyst-Coated Isotropic Colloids into "Active" Flexible Chains Enhances Their Diffusivity. *ACS Nano* **11**, 10025–10031 (2017).

33. Biswas, B., Misra, M., Singh Bisht, A., K. Kumar, S. & Kumaraswamy, G. Colloidal assembly by directional ice templating. *Soft Matter* **17**, 4098–4108 (2021).

34. Ni, S., Leemann, J., Buttinoni, I., Isa, L. & Wolf, H. Programmable colloidal molecules from sequential capillarity-assisted particle assembly. *Sci. Adv.* **2**, (2016).

35. Squires, T. M. & Bazant, M. Z. *Breaking Symmetries in Induced-Charge Electro-Osmosis and Electrophoresis*. *Journal of Fluid Mechanics* vol. 560 (2006).

36. Zhang, J., Grzybowski, B. A. & Granick, S. Janus Particle Synthesis, Assembly, and Application. *Langmuir* **33**, 6964–6977 (2017).

37. Biswas, B. *et al.* Rigidity Dictates Spontaneous Helix Formation of Thermoresponsive Colloidal Chains in Poor Solvent. *ACS Nano* **15**, 19702–19711 (2021).

38. Ni, S., Isa, L. & Wolf, H. Capillary assembly as a tool for the heterogeneous integration of micro- and nanoscale objects. *Soft Matter* **14**, 2978–2995 (2018).

39. Zhang, J. & Granick, S. Natural selection in the colloid world: Active chiral spirals. *Faraday Discuss.* **191**, 35–46 (2016).